\documentstyle[12pt]{article}
\def\be{\begin{equation}}
\def\te{\end{equation}}
\def\bea{\begin{eqnarray}}
\def\tea{\end{eqnarray}}
\def\nn{\nonumber}

\def\cD{{\cal D}}
\def\Tmn{T_{\mu\nu}(x,x')}
\def\Tc{T_{\mu\nu}^c(x,x')}
\def\cTmn{{\cal T}_{\mu\nu}(x,x')}
\def\cTc{{\cal T}_{\mu\nu}^c(x,x')}

\def\Kc{K^{\mu\nu}_c(x,x')}
\def\Gmn{G_{\mu\nu}}

\makeatletter                    
\@addtoreset{equation}{section}  
\makeatother                     

\begin{document}
\title{General Relativity as Geometro-Hydrodynamics
\thanks{This is an expanded version of an invited talk given at the
Second Sakharov International Symposium, Lebedev Physical Institute,
Moscow, May 20-24, 1996. The main ideas were discussed in two seminars given
by the author at the University  of Maryland, College Park in Fall, 1993}}
\author{B. L. Hu
\thanks{e-mail:hu@umdhep.umd.edu}\\
{\small Department of Physics, University of Maryland,
College Park, MD 20742, USA}}
\date{\small {\it (umdpp 96-114, July, 1996)}}
\maketitle

\begin{abstract}

In the spirit of Sakharov's `metric elasticity' proposal \cite{Sak}, we draw
a loose analogy between general relativity and the hydrodynamic state
of a quantum gas, and examine the various
conditions which underlie the transition from some candidate theory of
quantum gravity to general relativity, specifically, the long wavelength,
low energy (infrared) limits, the quantum to classical transition, the
discrete to continuum limit, and the emergence of a macroscopic collective state
from the microscopic consitituents and interactions of spacetime and fields.
In the `top-down' approach, we mention  how general relativity arises as
various limits are taken in some popular candidate theories of quantum gravity,
such as string theory, quantum geometry via the Ashtekar variables,
and simplicial quantum gravity. Our emphasis here is more on the `bottom-up'
approach, where one starts with the semiclassical theory of gravity and examines
how it is modified by graviton and quantum field excitations near and above
the Planck scale. We mention three aspects based on our recent findings:
1) Emergence of stochastic behavior of spacetime and matter fields depicted
by an Einstein-Langevin equation. The backreaction of quantum fields on the
classical background spacetime manifests as a fluctuation-dissipation relation.
2) Manifestation of stochastic behavior in effective theories below the
threshold arising from excitations above. The implication for general relativity
is that such Planckian effects, though exponentially suppressed, is in principle
detectable at sub-Planckian energies. 3) Decoherence of correlation histories
and quantum to classical transition. From  Gell-Mann and Hartle's observation
that the hydrodynamic variables which obey conservation laws are most readily
decohered, one can, in the spirit of Wheeler, view the conserved Bianchi
identity obeyed by the Einstein tensor as an indication that general relativity
is a hydrodynamic theory of geometry. Many outstanding issues surrounding the
transition to general relativity are of a nature similar to hydrodynamics and
mesoscopic physics.
\end{abstract}

\newpage
The abstract above outlines the philosophy and emphasis of our
current investigation. Details of the three main directions
are contained in the following references:

1) Einstein-Langevin equations and backreation in semiclassical gravity as
fluctuation-dissipation relation: \cite{HuPhysica,nfsg,fdrsc,HM3} (see also
\cite{CamVer})

2) Stochastic behavior of effective field theory across the threshold.
\cite{seft,sqed,sscg}

3) Decoherent correlation history and general relativity as geometro-
hydrodynamics. It is based on the ideas of Gell-Mann and Hartle
\cite{GelHar1,GelHar2}
(expanded in \cite{HLM,BruHal}) and Calzetta and Hu \cite{dch,cddn}
(detailed in \cite{CHP}).

Because of limited space, I will focus on only
the last of these three directions mentioned above.

Let me begin with the bigger picture of where we stand in understanding
the relation between the known classical theory of general relativity,
semiclassical gravity (where quantum fields are considered with classical
background spacetime) \cite{BirDav},
and the unknown but increasingly revealing domain
of quantum gravity.  Generally speaking, one can group all approaches
into two categories:

\subsection{`Top-down': How to reach the correct limits}

The transitional aspects mentioned in the abstract, i.e.,
quantum to classical transition,
low energy, long wavelength (infrared) limits,
discrete to continuum limit,
extended structure to point structure, and
micro/constituents versus  macro/collective states,
manifest in varying degrees of transparency in three leading types of candidate
theories of quantum gravity: the superstring theory \cite{string},
the loop representation of quantum geometry via spin connections \cite{loop},
and simplicial quantum gravity \cite{simpQG}. In string theory, a spin-two
particle is contained in the string excitations, and it is easy to see
the limit taken from an extended structure such as a p-brane to a point.
The Bekenstein-Hawking expression for the black hole entropy \cite{BekHaw}
is obtained as the tree level result of many quantum theories of gravity
\cite{tdbhent}. \footnote{Jacobson \cite{JacEqState}
has used the thermodynamic expression for black hole entropy to
show how  Einstein's equation can be derived as a thermodynamic equation of
state. The underlying philosophy of this view is similar to ours.}
But in the construction of a statistical mechanical entropy \cite{smbhent},
it is not so clear which of the many internal degrees of freedom of string
excitations contribute to the leading quantum correction term.
It is likely to be some collective
excitation state which dominate at low energy, like the collective degrees
of freedom depicting the vibrational and rotational motion of a nucleus, which,
though intuitively clear  from low energy, are not so easily
constructable from nucleon wavefunctions. The larger
problem of how the target space
(e.g., spacetime of 26 dimensions for bosonic string)
can be deduced from, or at least treated on the same footing as, the
world-volume of fundamental p-branes, still remains elusive.
In the quantum relativity approach using Ashtekar's spin connection and
Rovelli-Smolin's loop representation, the picture of a one-dimensional
quantum weave behaving like a polymer is evoked \cite{AshErice}.
When viewed at a larger scale the weaves appear to `knit' a higher
dimensional spacetime structure.
This is an interesting picture, but
how this collective process comes about --
i.e., how the physical spacetime becomes  a dynamically preferred entity
and an infrared stable structure -- remains to be explicated
(cf. protein-folding?). In simplicial quantum gravity \cite{simpQG},
the classical limit might be obtained more easily in some versions (e.g.,
in the Ponsano-Regge 6j calculus \cite{PonReg}, 
it is quite similar to the treatment of ordinary spin systems via
group-theoretical means, in place of the more involved considerations of
environment-induced decoherence
\cite{envdec}), but essential properties like diffeomorphism invariance
in the continuum limit are not guaranteed, such as in Regge calculus.
Dynamical triangulation procedure \cite{dyntriQG} was believed to work
nicely in these respects. But recently it was reported \cite{Kawai} that
a first order transition may arise which destroys the long wavelength niceties.
\footnote{I will not attempt to quote standard references in each of
these areas, as there are too many and fast-accumulating.
The reader can browse through this proceeding
and get a fairly good picture of the current status.}

Many structural aspects of these theories in their asymptotic regimes
(defined by the above-mentioned limits) near the Planck scale bear sufficient
resemblance to the physics in the atomic and nuclear scales
that I think it is useful to examine the underlying issues in the light of
these better-understood and well-tested theories. These include on the one
hand theories of `fundamental' interactions and constituents, such as
quantum electrodynamics (QED), quantum chromodynamics (QCD), supersymmetry
(SUSY), and grand unified theories (GUT) which are indeed what piloted
many of today's candidate theories of quantum gravity, and on the other hand
theories about how these interactions
and constituents manifest in a collective setting --
theories traditionally discussed in condensed matter physics using methods of
statistical mehanics and many-body theories.
These two aspects are not disjoint, but are interlinked in any realistic
description of nature (see \cite{HuHK,HuSpain}). They should be addressed
together in the search for a new theory describing matter and spacetime
at a deeper level.  The collective state description has not been emphasized
as much as the fundamental interaction description. We call  attention to its
relevance because especially in this stage of development of candidate
theories of quantum gravity, deducing their behavior and testing their
consequences at low energy  constitute an
important discriminant of their viability. Low energy particle
spectrum and black hole entropy are among the currently pursued topics.

Take, for example, the interesting observations related above,
that four-dimensional spacetime  is an apparent (as observed at low energy)
rather than a `real' (at Planck energy scale) entity (this is also highlighted
in Susskind's \cite{Sus,Jac} world as hologram  and 't Hooft's
\cite{tHo} view of the string theoretical basis of black hole dynamics and
thermodynamics, or the relation between the weave and the cloth,
in Ashtekar's program).
General relativity could be an emergent theory in some `macroscopic',
averaged sense at the low energy, long wavelength limit.
The fact that fundamental constituents manifest very different features at
lower energies is not such a surprising fact, they are encountered in
almost all levels of structure -- molecules from atoms, nuclei from quarks.
These are what I have referred to categorically as `collective states'.
How relevant and useful these variables or states are depend
critically on the scale and nature of
the physics one wants to probe. One cannot say that one is better than the
other without stipulating the range of energy in question, the nature of
the probe or the precision of the measurement. Just as thermodynamic variables
are powerful and
economical in the description of long wavelength processes, they are
completely useless at molecular scales. Even in molecular kinetic theory,
different variables (distribution and correlation functions) are needed for
different ranges of interactions. 
In treating the relation of quantum gravity to general relativity it is
useful to bear in mind these general features we learned from
more familiar processes.

Even when one is given the correct theory of the constituents, it is not
always an easy task to construct the appropriate collective variables
for the description of the relevant physics at a stipulated scale.
Note, e.g., how different the dynamical variables of the nuclear
collective models are from the independent particle model.
Not only are the derived structures different from their constituents, their
effective interactions can also be of a different nature.
There used to be a belief (myth) that once one has the fundamental theory it is
only a matter of details to work out an effective theory for its lower-energy
counterparts. Notice how nontrivial it is to deduce the nuclear force from
quark-gluon interactions, despite our firm knowledge that QCD is the progenitor
theory of nucleons and nuclear forces.
Also, no one has been clever enough to have derived, say,
elasticity from QED yet. Even if it
is possible to introduce the approximations to derive it, we know it is plain
foolish to carry out such a calculation, because at sufficiently low energy,
one can comfortably use the stress and strain variables for the description of
elasticity. (Little wonder quantum mechanics, let alone QED, is not a required
course in mechanical engineering!)

\subsection{`Bottom-Up': Tell-tale signs from low energy}

How the low energy behavior of a theory is related to its high energy behavior
(issues of effective decoupling and renormalizability naturally would arise
\cite{eft}), whether one can decipher traces
of its high energy interactions or  remnants of its high energy components,
have been the central task of physics since the discovery of atoms in the last
century and subatomic particles in this century to today's attack on unified
theories at ultrahigh energy.
The symmetry of the particles and interactions existing at low energies
are the raw data we need to construct (and rate the degree of success of)
a new unified theory. (Such is the central mission of e.g., string phenomenology.)
Some salient features of general relativity such as diffeomorphism invariance,
Minkowsky spacetime as a stable low energy limit, etc., are necessary
conditions for any quantum theory of gravity to meet.
Approaching Planck energy from below,
 the beautifully simple yet deep theory of
black hole thermodynamics \cite{BekHaw} based on semiclassical gravity,
has, and still is, providing
a checkpoint for viable quantum gravity theories -- the task now
is to obtain a microscopic (statistical mechanical) description of black hole
entropy beyond the tree-level.
I would like to supplement these two ongoing efforts (i.e., low energy particle
spectrum and black hole entropy) by proposing some new
directions, related to the existence of stochastic and fluctuation effects
at the cross-over regime.

\subsubsection{fluctuations and noise at the threshold}

An important feature of physics at the Planck scale depicted by semiclassical
gravity is the backreaction of  quantum effects
of particles and fields, such as vacuum polarization and particle creation,
on the classical gravitational spacetime. This is an essential step beyond
classical relativity for the linkage with quantum gravity.
For example, generalization to the $R + R^2$ theory of gravity
is a necessary product from the renormalization considerations of
quantum field theory in curved spacetimes. It should also be the low energy
form of string theory (plus dilaton and antisymmetric fields).
Backreaction demands more, in that the quantum matter field is solved
consistently with the classical gravitational field \cite{cosbkrn}.
The consistency requirement in a backreaction calculation brings in two
new aspects:\\
1) The classical gravitational field obeys a dynamics which contains a dissipation
component arising from the backreaction of particle creation in the quantum
field. The dissipation effect is in general nonlocal, as it is influenced by
particle creation not only occuring at one moment, but also integrated over the
entire history of this process \cite{pcB1}.\\
2) Creation of particles in the quantum matter field at the Planck energy
(which is responsible for the dissipative dynamics of the gravitational field)
can be depicted as a source which has both a deterministic and a stochastic
component. The first part is the averaged energy density of created particles,
which is known in previous treatments. The second part is new -- it measures
the difference of the amount of particles created in two neighboring histories
and is depicted by a nonlocal kernel, the correlator of a colored
noise \cite{nfsg,HM2}.  The dissipation and noise kernels are related by a
fluctuation-disspation relation which is a consequence of the unitarity
condition of the original closed (gravitation + matter field) system.

These processes are captured nicely by way of the influence functional \cite{if},
which is structurally equivalent to the Schwinger-Keldysh (closed-time-path or
in-in) effective action \cite{ctp}. It also elicits clearly the statistical
mechanical meaning of these quantum processes. (For a review of these
recent findings in the backreaction problems of semiclassical gravity, see, e.g.,
\cite{Banff}.)
The backreaction equation is in the form of a Langevin equation, which we call
the Einstein-Langevin equation \cite{HM3,fdrsc}.
The stochastic source term signifying Planckian quantum
field processes  integrates  to zero when one takes the ensemble average
with respect to the noise, reproducing the traditional semiclassical Einstein
equation. It is in the sense that we call general relativity
a mean field theory.
The Einstein-Langevin equation constitutes a new frontier for us to explore
possible phase transition and vacuum instability issues, which we believe many
of the `top-down' approaches would also encounter in this cross-over regime.

\subsubsection{stochastic behavior below the threshold}

What are the tell-tale signs for a low energy observer of the existence of
a high energy sector in the context of an effective field theory?
We wish to adopt an open system viewpoint to effective theory and explore
its statistical mechanical properties.
The question is to compare the difference between a
theory operative, i.e., giving an adequate description at low energies
(an open system, with the high energy sector acting as the environment)
to that of an exact low energy theory  by itself taken as a closed system.
We know that there are subtle differences between the two, arising from the 
backreaction of the heavy  on the light sector.
Though not obvious, the stochastic behavior associated with particle creation
above the threshold (which for gravitational processes is the Planck energy)
is related to the dissipative behavior of the background spacetime dynamics.
This was known for some time \cite{HuPhysica}. Schwinger's result \cite{Sch}
for pair production in a strong electromagnetic field 
is well-known. This effect at very low energy has however been ignored,
as it is usually regarded as background noise covered by very
soft photons. That such a noise carries information about the field at high
energy was pointed out only recently. 
Using a simple interacting field model, Calzetta and I found \cite{seft} 
that even at energy way below the threshold, stochastic effects, albeit at
extremely small amplitudes, can reveal some general (certainly not the 
specifics) properties of the high energy sector. We won't have
space to discuss the details, but refer the reader to \cite{seft} and earlier
references therein. We now turn to the decoherence aspects of quantum theories
and their classical limits to show why general relativity can be viewed
as the hydrodynamic limit of quantum gravity.

\section{Decoherence of Correlation History}

(This section contains a summary of work in \cite{dch,cddn}.)
The basic tenet of the consistent histories approach to quantum mechanics
\cite{conhis} is that quantum evolution may be
considered as the result of the coherent superposition of virtual fine-grained
histories, each carrying full information on the state of the system
at any given time. If we adopt the `natural' procedure of specifying a
fine-grained history by defining the value of the field $\Phi (x)$ at every
space time point, these field values being c numbers, then the quantum
mechanical amplitude for a given history is $\Psi [\Phi ]\sim e^{iS[\Phi ]}$,
where $S$ is the classical action evaluated at that particular history.
The virtual nature of these histories is manifested through the occurrence
of interference phenomena between pairs of histories. The strength of these
effects is measured by the `decoherence functional'

\begin{equation}
\label{popa}{\cal D}_F[\Phi ,\Phi ^{\prime}]\sim \Psi [\Phi ]\Psi [\Phi
^{\prime}]^*\sim e^{i(S[\Phi ]-S[\Phi ^{\prime}])}
\end{equation}
In reality, actual observations in a classical world correspond to
`coarse-grained' histories. 
A coarse-grained history is defined in terms of a `filter function' $\alpha $,
which determines which fine-grained histories belong
to the superposition, and their relative phases. \footnote{For example, we may
define a coarse-grained history of a system with two degrees of
freedom $x$ and $y$ by specifying the values $x_0(t)$ of $x$ at all times.
Then the filter function is $\alpha [x,y]=\prod_{t\in R}\delta (x(t)-x_0(t))$.}
The quantum mechanical amplitude for the coarse-grained history is defined as

\begin{equation}
\label{psiofalfa}\Psi [\alpha ]=\int~D\Phi~e^{iS}\alpha [\Phi ]
\end{equation}
where the information on the quantum state of the field is assumed to have
been included in the measure and/or the boundary conditions for the
functional integral. The decoherence functional for two coarse-grained
histories is  \cite{dch}

\begin{equation}
\label{dofalfa}{\cal D}_F[\alpha, \alpha ^{\prime}]=\int~D\Phi D\Phi'
e^{i(S[\Phi]-S[\Phi'])} \alpha [\Phi ]\alpha ^{\prime}[\Phi' ]^*
\end{equation}
In this path integral expression, the two histories $\Phi$ and $\Phi'$
are not independent; they assume identical values on a $t=T={\rm constant}$
surface in the far future. These are the same boundary
conditions satisfied by the histories on each branch of the time path in the
closed-time-path (CTP) formalism \cite{ctp}. There,
one considers the specified kernels as products of
fields defined on a closed time-path. As such, one may
define up to four different kernels $G^{ab}$ ($a, b = 1, 2$ are the
CTP indices) independently, to be
identified with the four different possible orderings of the
fields. If the kernels $G^{ab}$ can actually be
decomposed as products of c-number fields on the CTP, then
we associate to them the quantum amplitude for correlation histories
(of second order here) as

\begin{equation}
\Psi [G^{ab}]=\int~D\Phi^a~e^{iS}\prod_{x\gg x',ab}
\delta (\Phi^a(x)\Phi^b(x')-G^{ab}(x,x'))\label{psiofGab}
\end{equation}
where $S$ stands for the CTP classical action.
The path integral can be manipulated to yield \cite{dch}
 
\begin{equation}
\Psi [G^{ab}]\sim [{\rm Det}\{{\partial^2\Gamma\over
\partial G^{ab}\partial G^{cd}}\}]^{(1/2)}
e^{i\Gamma [G^{ab}]}\label{psiofG,2PI,CTP}
\end{equation}
where $\Gamma$ stands for the closed-time path two-particle irreducible 
(CTP 2PI) effective action. This last
expression can be analytically extended to more general propagator
quartets (and, indeed, even to kernels which do not satisfy the
relationships $G^{11}(x,x')=G^{21}(x,x')=G^{12*}(x,x')=G^{22*}(x,x')$
for $t\ge t'$, which follow from their interpretation as field
products).

Considering two histories associated with kernels $G(x,x')$ and $G'(x,x')$,
which can in turn be written as products of fields, we can construct
the decoherence functional for the second correlation order
as \cite{dch}

\begin{eqnarray}
\cD[G,G']&&=
\int~D\Phi D\Phi '~
e^{i(S[\Phi ]-S[\Phi '])}\nonumber\\
&&\prod_{x\gg x'}\delta ((\Phi (x)
\Phi (x')-G(x,x'))\delta ((\Phi '(x)
\Phi '(x')-G'^*(x,x'))
\label{DofGG'}
\end{eqnarray}

Using the  expression  Eq.(\ref{psiofGab}) for the quantum amplitude
associated with the most general binary correlation history, we can rewrite
Eq. (\ref{DofGG'}) as
 
\begin{equation}
\cD[G,G']=\int~DG^{12}~DG^{21}~\Psi [G^{11}=G,G^{22}=G'^*,G^{12},G^{21}]
\label{DofGG'2pi}
\end{equation}
 
In the spirit of our earlier remarks, we  can use the ansatz
Eq.(\ref{psiofG,2PI,CTP}) for the CTP quantum amplitude and
perform the integration by saddle point methods to obtain
 
\begin{equation}
\cD[G,G']\sim e^{i\Gamma [G^{11}=G,G^{22}=G'^*,G^{12}_0,G_0^{21}]}
\label{DofGG'CTP}
\end{equation}
where the Wightman functions are chosen such that
 
\begin{equation}
{\partial\Gamma\over\partial G^{12}_0}=
{\partial\Gamma\over\partial G^{21}_0}=0
\label{2pisaddle}
\end{equation}
for the given values of the Feynman and Dyson functions.

One can generalize this construction to higher correlation orders.
Indeed, in appealing to Haag's `reconstruction theorem',
(which states that the set of all expectation values of time - ordered
products of fields carries full information about the state of the
system \cite{Haag}) one can consider fine-grained histories as specified by the
given values of the irreducible time - ordered correlation functions. These
histories include those defined by the local value of the field, as those
where all irreducible Feynman functions vanish, but allow also more
general possibilities. Coarse grained histories will be specified by finite
sets of Feynman functions, and correspond to the truncated theories
described in detail in \cite{cddn}.

In particular, consider two histories defined by two sets of mean fields,
Feynman propagators and correlation functions up to $l$ particles, $\{\Phi
,G,C_3,...C_l\}$ and $\{\Phi ^{\prime },G^{\prime },C_3^{\prime
},...C_l^{\prime }\}$, respectively. Then the decoherence functional between
them is given by

\begin{equation}
\label{papa}{\cal D}_F[\{\Phi, G, C_3,...C_l\}, \{\Phi^{\prime}, G^{\prime},
C^{\prime}_3,...C^{\prime}_l\} ]\sim [prefactor] e^{i\Gamma_l}
\end{equation}
where $\Gamma_l$ is the $l$-loop CTP effective action evaluated at the
following history: (the prefactor is not important for our discussion)
a) Correlation functions on the `direct' branch are defined according to
the first history: $\Phi=\phi$, $G^{11}=G$, etc.
b) Those on the `return' branch are identified with the time-reverse of those
in the second history: $\Phi'=\phi'$, $G^{22}=(G^{\prime })^{*}$, etc.
c) All others are slaved to these.
(See \cite{cddn} for the meaning of `slaving')

The decoherence functional for correlation histories is a generalization of
the Griffith-Omnes and Gell-Mann-Hartle decoherence functional (\ref{popa})
between histories: it reduces to the latter when all
irreducible Green functions are chosen to vanish. It is consistent, in
the sense that further integration over the higher correlation functions
$\{C_{k+1},...C_l\}$, say, gives back the decoherence functional appropriate
to the $k$ loop theory in the saddle point approximation to the trace.

Decoherence means physically that the different coarse-grained histories
making up the full quantum evolution are individually realizable and may
thus be assigned definite probabilities in the classical sense.
Therefore, the quantum nature of the system will be shed to the degree of
accuracy afforded by the coarse-graining procedure, and the dynamics
is described by a  self-consistent, coupled set of equations
of a finite number of (nonlocal) c-number variables.

In this finite, truncated theory,  decoherence is associated with information
degradation and loss of full predictability \cite{GelHar2}.
The effective dynamics of the open system becomes dissipative and acquires
a stochastic component. The noise incurred
from the truncation of the Dyson-Schwinger hierarchy and the slaving of
higher correlation functions is called
`correlation noise'. We have shown \cite{cddn} that the relationship
between dissipation and noise is embodied in the fluctuation-dissipation
relation. In the following we shall show how this formalism
can be conveniently applied to the consideration of hydrodynamic variables.

\section{Decoherence and Classicality in  Hydrodynamic Variables}

\subsection{Energy-Momentum Tensor of Scalar Fields}

(Discussion in this subsection is based on work currently in progress
with Calzetta and Paz \cite{CHP}.)
As an example of the correlation history formalism, let us consider operators
defined in terms of quadratic products of the fields
\be
T_{\mu\nu} (x, x') = {\cal T}_{\mu\nu} \Phi(x)\Phi(x')
\te
 where
 ${\cal T}$ deotes some differential operator on fields defined at two points.
For the energy momentum operator, the usual expression $T_{\mu\nu} (x)$
is the coincidence limit of $\cTmn$, when the two points
$x,x'$ are identified.
Let the diagonal components of the energy-momentum tensor be the energy
density $\rho (x)$ and the mementum density $p$, then
\be
\rho(x) = Lim_{x\rightarrow x'} {\cal T}_{00} (x,x') G(x,x')
\te
where the Green function $G(x,x')$ is  one of the two point functions 
constructed from the product of $\Phi(x)$ and $\Phi(x')$.
We want to show that the decoherence functional ${\cal D}$
of correlation histories peaks for operators of fields  ${\cal T}$
which obey a conservation law
\be
{T_{\mu\nu}}^{;\nu} = 0
\te
Physically this means that quantities which are conserved would be most likely
to be decohered and acquire a classical attribute.
Using the expression we derived in the previous section for correlation
histories of the second order, we can write down the decoherence functional
for ${\cal T}$ to be

\begin{eqnarray}
\cD[T_{\mu\nu},T'_{\mu\nu}]&&=
\int~D\Phi D\Phi '~e^{i(S[\Phi ]-S[\Phi '])}
\prod_{x\gg x'}\delta (\Tmn-  \cTmn    \Phi (x)\Phi (x')) \nn\\
&&                 \delta (\Tmn'- \cTmn'^* \Phi'(x)\Phi'(x'))
\label{DofTT'1}
\end{eqnarray}
Introducing the integral representation of the delta functional and using
the CTP indices $a,b,c= 1,2$ to include both the forward and backward
integrations, we can write this expression as,

\begin{equation}
\cD[T_{\mu\nu},T'_{\mu\nu}]=
\int~D\Phi^a \int~D\Kc e^{iS[\Phi^a ]}
~e^{i\Kc [\Tc - \cTc \Phi^a(x)\Phi^b(x')] }
\label{DofTT'2}
\end{equation}
For an action of the form $S [\Phi^a] = \Phi^a \Delta_{ab} \Phi^b $, we get

\begin{eqnarray}
\cD[T_{\mu\nu},T'_{\mu\nu}]&&=
 \int~D \Kc \int~D\Phi^a ~e^{i\Phi^a[\Delta + \Kc \cTc]_{ab} \Phi^b}
~e^{i\Kc \Tc} \nonumber\\
&& \sim e^{i\Gamma [\Tmn, \Tmn']}
\label{DofTT'}
\end{eqnarray}

This is the framework which one could use to validate the statement above,
\footnote{This is not a straightforward proof and we have not succeeded in
this way. An alternative approach is to rely on a kinetic theory proof of
how hydrodynamics is derived from the BBGKY hierarchy, i.e., by taking
the long wavelength and collision-dominated limit of equations of motion
from the nPI effective action in the Schwinger-Dyson hierarchy with
conservation laws imposed on the hydrodynamical variables.}
i.e., that by virtue of the conservation law
obeyed by $\Tmn$, the decoherence functional peaks with respect to the
hydrodynamic variables ($\rho, p)$. These variables are most readily decohered
and have the greatest chance of becoming classical.
(In the words of Gell-Mann and Hartle, they have the greatest
`inertia'.) Using the relation between the decoherence and influcence functional
as shown in \cite{nfsg}, one can further evaluate the noise and fluctuations
around the classical trajectory defined by the equations of motion derived
from the influence functional in a stochastic form.

\subsection{Hydrodynamics of Geometry: Bianchi Identities and Classicality}

Using the argument above, one can ask the same question about the properties
of geometric quantities such as the Einstein tensor $\Gmn$,  which
shares similar properties as $T_{\mu\nu}$. Its conservation by virtue of
the Bianchi identity ${\Gmn}^{;\nu}=0$ would likewise make the decoherence
functional between various quantum geometries peak, and be placed as the
hydrodynamic variables of geometries. It is in this sense (amongst
other considerations) that we can regard general relativity (Einstein's equations
governing the dynamics of geometry -- or geometrodynamics \cite{WheelerGMD})
as gemetro-hydrodynamics. Wheeler has long since emphasized the significance
of the Bianchi identity in relation to the basic laws of physics QED, QCD.
The geometric meaning of it amounts to the almost naive statement of
``the boundary of a boundary is zero" \cite{MTW}. This is what constitutes
his ``Austerity Principle" which governs all laws of physics \cite{WheelerAus}.
Here, the significance of the Bianchi identity is in providing a criterion
for picking out the variables with the largest `inertia' which can decohere
most readily \cite{GelHar2} -- the  `collective' or `hydrodynamic' variables
of geometry. These are the variables
which enter into the classical laws governing the dynamics of geometry, i.e.,
general relativity.

In considering  classical geometry as hydrodynamic variables, a natural question
evoked from this view is: What would correspond to quantum geometry? We are
allowing for the possibility that geometry can still be a valid concept
in a progenitor theory for a quantum description of spacetime, although
at small distances, most likely it would have lost its smoothness property.
It could involve non-commutative characters \cite{Connes} and
non-trivial topologies, of which string theory and spin network \cite{Penrose}
are capable of addressing. New ideas like quantum topology and logic
have also been proposed, and related to the Gell-Mann
Hartle decoherent history conceptual scheme \cite{Isham}.

In light of the connection between the quantum and the classical provided
by the decoherence conceptual scheme, one should view the classical world
as an emergent structure, the conservation laws are what give them the
relative stability and persistence characters.
In addition to recovering the classical
equations of motion from the extremum of the decoherence functional,
one can also get a measure of how stable it is with respect to
the influence of noise and fluctuations from the `environment'
(in the correlation approach, it is provided by the higher-order correlation
functions). Fluctuations and dissipation in the dynamics of the effective
system provide us with a window to look beyond the macro-classical world into
its micro-quantum basis.  Even starting from pure thermodynamics,
without the knowledge of statistical mechanics based on  quantum levels, one
can obtain some important properties of the system from its thermodynamic
functions.\footnote{One could gain quite a bit of information from
studying the fluctuations of a system.
This was the way critical phenomena was studied in the
Fifties and Sixties -- e.g., the heat capacity at constant volume measures the
thermodynamic stability of the canonical ensemble, the compressibility
at constant pressure measures the stability of the grand canonical ensemble.}
Many important thermodynamic properties of a system can be obtained in
the linear response regime, by considering how the system reacts to small
stimulus from an external force. This approach is useful for studying possible
phase transitions at Planck energy in the early universe
\cite{nfsg,fdrsc,HM3,CamVer} or to obtain a statistical mechanics generalization
of the Bekenstein-Hawking relation for black hole entropy \cite{tdbhent},
problems we are currently investigating.
Let us return to some more general questions generated from this
new way of viewing general relativity as low energy physics.

\section{Implications of this viewpoint: low energy collective state physics
and beyond}

Suppose one takes this viewpoint seriously, what are the possible implications?
We can make a few observations here.

\subsection{Quantizing metric may yield only phonon physics}

First, the laws governing these collective variables are classical,
macroscopic laws. It may not make full sense
to assume that by quantizing these variables directly
one would get the micro-quantum basis of the macro-classical theory,
as has been the dominant view in quantum gravity.
Just like the energy density $\rho$ and momentum densities $p$ in $\Tmn$,
which are the hydrodynamic variables of a matter field,
quantization should only be performed on
the microscopic fields $\Phi (x)$  from which they are constructed.
If one did so for the metric or the connection variables, one would
get the quantum excitations of geometry in the nature of phonons in relation to
atoms (or other quantum collective excitations in condensed matter physics).
That may be the next order of probe for us, and may yield some new physics,
but it is still very remote from seeing the nucleon structure
in the solid lattice or the attributes of quantum electrodynamics.
In the analogy we mentioned above, quantum elasticity
tells us nothing about QED!

Second -- and this is perhaps the more interesting aspect -- assuming that
the metric and connections are the collective variables, from the way they
are constructed,  what can one say about their microscopic, quantum basis?
Historically this question was asked repeatedly when one probes from low to
high energy scales, trying to decipher the microscopic consituents and laws of
interactions from macroscopic phenomena. This is like going from phonons to
the structure of atoms, from nuclear rotational spectrum to nucleon strucuture
-- not an easy question to answer. But there are nevertheless ways to guide us,
e.g., in terms of the tell-tale signs mentioned earlier.
In the above analogies, recall that atomic spectroscopy reveals many
properties about the electron-electron and electron-nucleus interactions,
low temperature anomalous behavior of specific heat reveals the quantum
properties of electrons, the intermediate boson model bridges the symmetry
of the collective modes with that of the independent nucleons.
To address questions like this, one needs to proceed from both ends:
One needs to postulate a theory of the microscopic structure, and work out
its collective states at large scale and low energies.
One also needs to comb through the consequences of the known low energy theory,
paying attention to subtle inconsistencies or mistakenly ignored trace effects
from higher energy processes. Indeed, this is what is going on today,
with string theory as the micro theory, and semiclassical gravity as its
low energy limit.
The viewpoint we are proposing would suggest focusing on collective
states (solitons?) of  excitations of the fundamental string on the one hand
and a detailed study of the possible new phenomena in quantum field theory
in curved spacetime, such as flucutations and phase transitions around the
Planck energy, and quantum corrections to the black hole entropy.

\subsection{Common features of collective states built from different constituents}

As mentioned above, there are two almost orthorgonal perspectives in
depicting the structure and properties of matter. One is by way of its
constituents and interactions, the other according to its collective behaviors.
The former is the well-known and well-trodden path of discovery of QED, QCD,
etc. If we regard this chain of QED - QCD - GUT - QG as a vertical
progression depicting the hierarchy of basic constituents, there is also a
horizontal progression in terms of the stochastic - statistical
- kinetic - thermodynamic/hydrodynamic depiction of the collective states.
It should not surprise us that there exist similarites between matters in
the same collective state (e.g., hydrodynamics) but made from different
constituents.
Macroscopic behavior of electron plasmas are similar in many respects to the
quark-gluon plasma. Indeed, one talks about magneto-hydrodynamics from Maxwell's
theory as well as magneto-chromo hydrodynamics from QCD. In this long wavelength,
collision-dominated regime, they can both be depicted by the hydrodynamics of
fluid elements, which are governed simply by Newtonian mechanics.
The underlying
micro-theories are different, but the hydrodynamic states of these constituents
are similar. Here we are proposing that general relativity could be viewed in
a similar way, i.e, it is the geometro-hydrodynamics from some candidate theory
of quantum gravity. It is an effective theory
in the way that nuclear physics is with regard to QCD, and atomic physics is
with regard to QED. They are all low energy collective states
of a more fundamental set of laws. The macroscopic, hydrodynamic equations
and their conservation laws like the Naviers-Stoke and the continuity
equations of hydrodynamics are all based on dynamical and conservation laws of
microphysics (e.g., Newtonian mechanics), but when expressed in terms of the
appropriate collective variables, they can take on particularly simple and
telling forms. Thermodynamic variables like temperature, entropy, etc.
(think black hole analogy -- mass, surface area) are derived quantities
with their specific laws (three laws)
traceable via the rules of statistical mechanics (of Gibbs and Boltzmann)
to the laws of quantum mechanics.  Rules of statistical mechancis are important
when we probe into a deeper layer of structure from known low energy theories
such as semiclassical gravity: we need to know how to disentangle the collective
states to sort out how the microphysics works. \footnote{Savour the importance
of, say, coming up with a statistical mechanical definition of temperature
in a canonical ensemble as the rate of change of the accessible states of a
system in contact with a heat reservoir with respect to changes in energy, and
we can appreciate the importance of Gibbs' work in relation to quantum physics.}
It is hard to imagine how a complete theory of microphysics can be attained
without going through this step.

One comment about symmetry laws: If we view general relativity as a
hydrodynamic theory in the same sense as the nuclear rotational and vibration
states in the collective or liquid drop model, we can see that the symmetries
of rotational and vibrational motion are useful description of the large scale
motion of a nucleus, but has no place in the fundamental symmetries of nucleons,
much less their constituents, the quarks and gluons. In this sense one could
also question the neccesity and legitimacy of familar concepts like Lorentz
invariance and diffeomorphism invariance in the more fundamental theory. It
should not surprise us if they no longer hold for trans-Planckian physics.

\subsection{Hydrodynamic fluctuations applied to black holes and cosmology}

A problem where this analogy with collective models may prove useful
is that of black hole entropy.
If we view the classical expression for black hole entropy to be a
hydrodynamic limit, and the corrections to it as arising from hydrodynamic
fluctuations, one could use linear response theory to approach conditions
near thermodynamic equilibrium and construct a non-equilibrium theory of
black hole thermodynamics.
\footnote{Black hole backreaction problem has been studied by many authors
before, notably by York \cite{York}, Anderson and Hiscock \cite{AndHis} and
their collaborators. We are taking a non-equilibrium statistical field theory
approach.
We aim to get the fluctuations of the energy momentum tensor of a quantum
field in a perturbed Schwarzschild spacetime \cite{HPR},
examine how they might induce dissipations of the event horizon and deduce a
susceptibility function of the black hole \cite{HRS}.
This would realize the proposal of Sciama that a black hole
in equilibrium with its Hawking radiation can be depicted in terms of a
fluctuation-dissipation relation \cite{CanSci}. (See also \cite{Mottola,Pavon})}
It also seems to us that many current attempts to
deduce the quantum corrections of black hole entropy
from the micro-quantum theory of strings
could be missing one step. This is like the correpondance between results predicted
from the independent particle (nucleon) model (where one can construct
the shell structures), and that from the liquid drop model (where one can
construct the collective motions) -- a gap exists which cannot easily be
filled by simple extensions of either models operative in their
respective domains of validity. This involves going from  the individual
nucleon wavefunctions to the collective states of a nucleus.
It is likely that only some
appropriate combinations of fundamental string excitation modes
which survive in the long wavelength limit can contribute to the excitations
of the collective variables (area and surface gravity of black hole)
which enter in the (semiclassical gravity) black hole entropy.

Another example where viewing the classical GR as a hydrodynamic limit would
help us orient our conceptual inquiries is in semiclassical cosmology
at the Planckian scale. As mentioned above, the abundance of particle creation
at that energy makes the consideration of their backreactions important,
and to accomodate the fluctuations of the field and the geometry
the conventional semiclassical Einstein equation needs be generalized to
the form of an Einstein-Langevin equation. In the view proposed here,
this would correspond to the hydrodynamic fluctuations of spacetime dynamics
as induced by these quantum field processes.
One could study the behavior of metric and field
fluctuations with this Langevin equation in a way similar to that of
critical dynamics for fluids and condensed matter.

\subsection{Planck scale resonance states}

Finally, following the progression from hydrodynamics to kinetic theory and
quantum micro- dynamics, one may ask if there could exist quasistable structures
at energy scales slightly higher than (or observation scales finer than)
the semiclassical scale. Assuming that string theory is the next level micro-theory,
does there exist quasi-stable structures between that and general relativity?
This is like the existence of resonance states (as quasi-stable particles)
beyond the stable compounds of quarks (baryons) or quark-antiquarks (mesons).
Viewed in the conceptual framework of kinetic theory, there could exist such
states, if the interparticle reaction times (collision and exchange)
and their characteristic dynamics (diffusion and dissipation) become
commensurate at some energy scale. (Turbulance in the nonlinear regime
are abundant in these intermediate states). In the framework of decoherent
history discussed above, it could also provide metastable quasiclassical
structures. It would be interesting to find out if such structures can
in principle exist around the Planck scale. This question is stimulated
by the hydrodynamic viewpoint, but the resolution would probably have to
come from a combination of efforts from both the top-down and the bottom-up
approaches. Deductions from high energy string theories would also
need the guidance from the different collective states which can exist
in the low energy physics
of general relativity and semiclassical gravity.

In closing, we note that progress of physics --
the probing of the structure and dynamics of matter
and spacetime -- has always moved in the direction from  low to high energies.
One needs to pay attention to the seemingly obvious facts at low energies
and probe into any discrepancy or subtlties not usually observed to find
hints to the deeper structures.
By examining how certain common characteristics of all successful low energy
theories  (here, we only discuss the hydrodynamic and thermodynamic aspects)
may recur in a new theory at a higher energy, and how they differ, we can
perhaps learn to
ask the right questions and focus on some hitherto neglected aspects.\\



\noindent {\bf Acknowledgement}
The work on correlation histories (Sec. III) is done with Esteban Calzetta,
and its application to the derivation of equation of motion from
conserved quantities like the  energy-momentum tensor (Sec. IV)
is on-going work with Esteban Calzetta and Juan Pablo Paz.
I thank them warmly for many years of close exchanges and collaborations.
Appreciation goes to Ted Jacobson for useful comments on a preliminary draft.
Research is supported in part by the National Science Foundation 
under grant PHY94-21849.



\begin{thebibliography}{999}

\bibitem {Sak}
A. D. Sakharov, ``Vacuum Quantum Fluctuations in Curved Space and the Theory
of Gravitation'' Doklady Akad. Nauk S. S. R. 177, 70-71 (1987)
[Sov. Phys. - Doklady 12, 1040-1041 (1968)]. See also S. L. Adler, Rev. Mod. 
Phys. 54, 729 (1982)

\bibitem{HuPhysica} B. L. Hu, Physica A158, 399 (1989)

\bibitem{nfsg}
E. Calzetta and B. L. Hu, Phys. Rev. {\bf D49 }, 6636 (1994)

\bibitem{fdrsc}
B. L. Hu and S. Sinha, Phys. Rev. {\bf D51}, 1587 (1995).

\bibitem {HM3}
B. L. Hu and A. Matacz,
Phys. Rev. D {\bf 51}, 1577 (1995)

\bibitem{CamVer}
A. Campos and E. Verdaguer, Phys. Rev. D53, 1927 (1996)

\bibitem{seft}
E. Calzetta and B. L. Hu, ``Stochastic Behavior of Effective Field Theories
Across Threshold" hep-th/9603164

\bibitem{sqed}
E. Calzetta, B. L. Hu and F. D. Mazzitelli, "Stochastic Effects in
Quantum Electrodynamics" (in preparation)

\bibitem{sscg}
E. Calzetta, B. L. Hu and F. D. Mazzitelli, "Stochastic Effects in
General Relativity from Planckian Processes" (in preparation)

\bibitem {GelHar1}
M. Gell-Mann and J. B. Hartle, in {\it Complexity, Entropy and the Physics
of Information}, ed. by W. H. Zurek (Addison-Wesley, Reading, 1990)

\bibitem{GelHar2}
M. Gell-Mann and J. B. Hartle, Phys. Rev. {\bf D47}, 3345 (1993)

\bibitem{HLM}
J. B. Hartle, R. Laflamme and D. Marolf, Phys. Rev. D51, 7007 (1995)

\bibitem{BruHal}
T. A. Brun and J. J. Halliwell, Phys. Rev. D53 (1996)

\bibitem{dch}  E. Calzetta and B. L. Hu, ``Decoherence of Correlation
Histories'' in {\it Directions in General Relativity, Vol II: Brill
Festschrift}, eds B. L. Hu and T. A. Jacobson (Cambridge University Press,
Cambridge, 1993) gr-qc/9302013

\bibitem {cddn} E. Calzetta and B. L. Hu, ``Correlations, Decoherence,
Disspation and Noise in Quantum Field Theory'', in {\it Heat Kernel
Techniques and Quantum Gravity}, ed. S. Fulling (Texas A\& M Press, College
Station 1995). hep-th/9501040


\bibitem{CHP}
E. Calzetta, B. L. Hu and J. P. Paz, ``Einstein's Equation from the Decoherence
of Correlation Histories" (in preparation)

\bibitem{BirDav}  N. D. Birrell and P. C. W. Davies, {\it Quantum Fields in
Curved Spaces} (Cambridge University Press, Cambridge, 1981).

\bibitem{string}
M. B. Green, J. H. Schwarz and E. Witten,
{\it Superstring Theory} (Cambridge University, Cambridge, 1990).
E. Witten, Physics Today 49, 24 (1996)

\bibitem{loop}
A. Ashtekar, {\it Lectures on Non-Perturbative Canonical Gravity}
(World Scientific, Singapore, 1991);
A. Ashtekar and J. Lewandowski, in {\it Knot Theory and Quantum Gravity}
ed. J. Baez (Oxford University, London, 1995).

\bibitem{simpQG}
T. Regge, Nuovo Cimento 19, 558 (1961).
For  recent work, see, e.g.,
J. B. Hartle, J. Math. Phys. 26, 804 (1985); 27, 287 (1986); 30, 452 (1989).
H. W. Hamber, in {\it Critical Phenomena, Random Systems, Gauge Theories}
      1984 Les Houches Summer School, eds K. Osterwalder and R. Stora
      (North Holland, Amsterdam, 1986).
H. W. Hamber, Nucl. Phys. B (Proc. Suppl) 20, 728 (1991); 25A, 150 (1992);
     Phys. Rev. D45, 507 (1992); Nucl. Phys. B400, 347 (1993);
R. M. Williams and P. A. Tucker, Class. Quant. Grav. 9, 1409 (1992).
H. W. Hamber and R. M. Williams, Phys. Rev. D47, 510 (1993);
      Nucl. Phys. 415, 463 (1994)

\bibitem {BekHaw}
J. D. Bekenstein, Phys. Rev. D7, 1333 (1973).
S. W. Hawking, Commun. Math. Phys. {\bf 43}, 199 (1975).

\bibitem {tdbhent}
R. M. Wald, Phys. Rev. D48, R3427 (1993).
T. Jacobson, G. Kang, and R. Myers, Phys. Rev. D49, 6587 (1994).
M. Banados, C. Teitelboim and J. Zanelli, Phys. Rev. Lett. 72, 957 (1994).
J. D. Bekenstein, Review Talk at the Seventh Marcell Grossmann
Meeting, Stanford, 1994.  gr-qc/9409015.
D. M. Page, ``Black Hole Information", in {\it Proc. 5th Canadian Conference
on General Relativity and Relativistic Astrophysics} eds. R. B.
Mann and R. G. McLenaghan (World Scientific, Singapore, 1994) hep-th/9305040.
V. P. Frolov, D. V. Fursaev and A. I. Zelnikov, ``Black Hole Entropy:
Off-Shell vs On-Shell" hep-th/9512184.

\bibitem{JacEqState}
T. Jacobson, Phys. Rev. Lett.  (1995)

\bibitem {smbhent}
L. Susskind and J. Uglam, Phys. Rev. D50, 2700 (1994);
J. D. Bekenstein and V. F. Mukhanov, Phys. Lett. B (1995);
G. T. Horowitz, ``The Origin of Black Hole Entropy in String Theory"
in the Proceedings of the Pacific Conference on Gravitation and Cosmology,
Seoul, Korea, Feb, 1996. gr-qc/9604051

\bibitem {AshErice}
A. Ashtekhar, in {\it String Gravity and Physics at the Planck Energy Scale}
eds. N. Sanchez and A. Zichichi (Kluwer, Dordrecht, 1996)

\bibitem {PonReg}
G. Ponzano and T. Regge, "Semiclassical limit of Racah coefficients"
in {\it Spectroscopic and Group Theoretical Methods in Physics}
ed. F. Bloch (North Holland, Amsterdam, 1968)
J. Iwasaki, ``A reformulation of the Ponzano- Regge quantum gravity models
in terms of surfaces" (1994)
J. W. Barrett and T. J. Foxon, Class. Quant. Grav. 11, 543 (1994).
J. W. Barrett, ``Quantum Gravity as topological quantum field theory" (1995)

\bibitem{envdec}  W. H. Zurek, Phys. Rev. D24, 1516 (1981); D26, 1862
(1982); in {\it Frontiers of Nonequilibrium Statistical Physics}, ed. G. T.
Moore and M. O. Scully (Plenum, N. Y., 1986); Physics Today {\bf 44}, 36
(1991); E. Joos and H. D. Zeh, Z. Phys. B59, 223 (1985); A. O. Caldeira and
A. J. Leggett, Phys. Rev. {\bf A 31}, 1059 (1985); W. G. Unruh and W. H.
Zurek, Phys. Rev. D40, 1071 (1989);  B. L. Hu, J. P. Paz and Y. Zhang,
Phys. Rev. {\bf D45}, 2843 (1992);
W. H. Zurek, Prog. Theor. Phys. 89, 281 (1993).

\bibitem {dyntriQG}
J. Ambj\o rn, B. Durhuus and J. Fr\"olich, Nucl. Phys. B257 [FS14], 433 (1985);
B275 [FS17], 161 (1986);
J. Ambj\o rn, B. Durhuus, J. Fr\"olich and P. Orland, Nucl. Phys. B270 [FS16],
457 (1986);
F. David, Nucl. Phys. B257 [FS14], 45, 543 (1985);  Phys. Lett. 159B, 303 (1985)
V. A. Kazakov, Phys. Lett. 150B, 282 (1985);
V. A. Kazakov, I. K. Kostov and A. A. Migdal, Phys. Lett. 157B, 295 (1985);
E. Brezin and V. A. Kazakov, Phys. Lett. 236B, 144 (1990)
D. J. Gross and A. A. Migdal, Phys. Rev. Lett. 64, 717 (1990);
   Nucl. Phys. B340, 333 (1990)
M. R. Douglas and S. H. Shenkar, Nucl. Phys. B335, 635 (1990).
J. Ambj\o rn, B. Durhuus and T. Jonsson, Mod. Phys. Lett. A6, 1133 (1991);
M. E. Agishtein and A. A. Migdal, Mod. Phys. Lett. A6, 1863 (1991)

For recent reviews, see, e.g., {\it Statistical Mechanics of Membranes
and Surfaces} eds D. Nelson et al (World Scientific, Singapore, 1989);
{\it Two-Dimensional Quantum Gravity and Random Surfaces} eds D. J. Gross,
T. Piran and S. Weinberg (World Scientific, Singapore, 1992).

\bibitem {Kawai}
See contributions by Marinari, Kazakov and Ambjorn in this Proceeding.

\bibitem{HuHK}
B. L. Hu, ``Cosmology as `Condensed Matter' Physics'' in Proc. Third  Asia
Pacific Physics Conference,
ed. Y. W. Chan et al (World Scientific, Singapore, 1988) Vol. 1, p. 301.
gr-qc/9511076

\bibitem{HuSpain}  B. L. Hu, ``Fluctuation, Dissipation and Irreversibility
in Cosmology'' in {\it The Physical Origin of Time-Asymmetry}, Huelva,
Spain, 1991 eds. J. J. Halliwell, J. Perez-Mercader and W. H. Zurek
(Cambridge University Press, Cambridge, 1994).

\bibitem {Sus}
L. Susskind, J. Math. Phys. 36, 6377 (1995)

\bibitem {Jac}
S. Corley and T. Jacobson, Phys. Rev. D53, R6720 (1996)

\bibitem {tHo}
G. 't Hooft,  ``Quantization of Point Particles in 2+1 Dimensional Gravity
and Spacetime Discreteness" gr-qc/9601014

\bibitem{eft}  T. Appelquist and J. Carazzone, Phys. Rev. {\bf D11}, 2856
(1975) S. Weinberg, Phys. Lett. {\bf 83B}, 339 (1979); B. Ovrut and H. J.
Schnitzer, Phys. Rev. {\bf D21}, 3369 (1980); {\bf D22}, 2518 (1980) L.
Hall, {\sl Nucl. Phys. }{\bf B178}, 75 (1981). P. Lapage, in {\it From
Actions to Answers}, Proc. 1989 Adv. Inst. in Elementary Particle Physics,
eds T. DeGrand and D. Toussaint (World Scientific, Singapore, 1990) p. 483.
S. Weinberg, {\it The Quantum Theory of Fields} (Vol. 1)
(Cambridge University Press, Cambridge, 1995).

\bibitem{cosbkrn}
Ya. Zel'dovich and A. Starobinsky, Zh. Eksp. Teor. Fiz {\bf 61}, 2161 (1971)
[Sov. Phys.- JETP {\bf 34}, 1159 (1971)]
L. Grishchuk, Ann. N. Y. Acad. Sci. 302, 439 (1976).
B. L. Hu and L. Parker, Phys. Lett. 63A, 217 (1977).
B. L. Hu  and L. Parker, Phys. Rev. {\bf D17}, 933 (1978).
F. V. Fischetti, J. B. Hartle and B. L. Hu, Phys. Rev. {\bf D20}, 1757 (1979).
J. B. Hartle and B. L. Hu, Phys. Rev. {\bf D20}, 1772 (1979).
{\bf 21}, 2756 (1980).
J. B. Hartle, Phys. Rev. D23, 2121 (1981).
P. A. Anderson, Phys. Rev. D28, 271 (1983); D29, 615 (1984).

\bibitem {pcB1}
B. L. Hu  and L. Parker, Phys. Rev. {\bf D17}, 933 (1978).
J. B. Hartle and B. L. Hu, Phys. Rev. {\bf D20}, 1772 (1979);
E. Calzetta and B. L. Hu, Phys. Rev. {\bf D35}, 495 (1987).

\bibitem {HM2}
B. L. Hu and A. Matacz, Phys. Rev. D49, 6612 (1994)

\bibitem{if}
R. Feynman and F. Vernon, Ann. Phys. (NY) {\bf 24}, 118 (1963).
R. Feynman and A. Hibbs, {\it Quantum Mechanics and Path Integrals},
(McGraw - Hill, New York, 1965).
A. O. Caldeira and A. J. Leggett, Physica {\bf 121A}, 587 (1983).
H. Grabert, P. Schramm and G. L. Ingold, Phys. Rep. {\bf 168}, 115
(1988).
B. L. Hu, J. P. Paz and Y. Zhang, Phys. Rev. {\bf D45}, 2843 (1992);
{\bf D47}, 1576 (1993)

\bibitem{ctp}
J. Schwinger, J. Math. Phys. {\bf 2} (1961) 407;
P. M. Bakshi and K. T. Mahanthappa,
J. Math. Phys. 4, 1 (1963), 4, 12 (1963);
L. V. Keldysh, Zh. Eksp. Teor. Fiz. {\bf 47 }, 1515 (1964)
[Engl. trans. Sov. Phys. JEPT {\bf 20}, 1018 (1965)].
G. Zhou, Z. Su, B. Hao and L. Yu, Phys. Rep. {\bf 118}, 1 (1985);
Z. Su, L. Y. Chen, X. Yu and K. Chou, Phys. Rev. {\bf B37}, 9810
(1988).
B. S. DeWitt, in {\it Quantum Concepts in Space and Time},
ed. R. Penrose and C. J. Isham (Claredon Press, Oxford, 1986);
R. D. Jordan, Phys. Rev. D33, 44 (1986).
E. Calzetta and B. L. Hu, Phys. Rev. {\bf D35}, 495 (1987).

\bibitem {Banff}
B. L. Hu, ``Quantum Statistical Fields in Gravitation and
Cosmology'' in {\it Proc. Third International Workshop on Thermal
Field Theory and Applications}, eds. R. Kobes and G. Kunstatter (World
Scientific, Singapore, 1994) gr-qc/9403061

\bibitem {Sch}
J. Schwinger, Phys. Rev. 82, 664 (1951).

\bibitem{conhis}  
R. B. Griffiths, J. Stat. Phys. {\bf 36}, 219 (1984).
R. Omn\'es, J. Stat Phys. {\bf 53}, 893, 933, 957 (1988); Ann. Phys. (NY) 
{\bf 201}, 354 (1990); Rev. Mod. Phys. {\bf 64}, 339 (1992); {\it The
Interpretation of Quantum Mechanics} (Princeton UP, Princeton, 1994).
M. Gell-Mann and J. B. Hartle, in {\it Complexity, Entropy and the Physics
of Information}, ed. by W. H. Zurek (Addison-Wesley, Reading, 1990); 
Phys. Rev. {\bf D47}, 3345 (1993) ;
J. B. Hartle, ``Quantum Mechanics of Closed
Systems'' in {\it Directions in General Relativity} Vol. 1, eds B. L. Hu,
M. P. Ryan and C. V. Vishveswara (Cambridge Univ., Cambridge, 1993).
H. F. Dowker and J. J. Halliwell, Phys. Rev. {\bf D46}, 1580 (1992); 
T. Brun, Phys. Rev. {\bf D47}, 3383 (1993);
J. P. Paz and W. H. Zurek, Phys. Rev. {\bf D48} 2728 (1993);
J. Twamley, Phys. Rev. {\bf D48}, 5730 (1993)
H. F. Dowker and G. Gauntlett, J. Stat. Phys. (1996)

\bibitem{Haag}  R. Haag, {\it Local Quantum Physics} (Springer, Berlin, 1992).

\bibitem {WheelerGMD}
J. A. Wheeler, {\it Geometrodynamics} (Academic Press, New York, 1962)

\bibitem {MTW}
C. W. Misner, K. S. Thorne and J. A. Wheeler, {\it Gravitation}
(Freeman, San Francisco, 1973).

\bibitem {WheelerAus}
J. A. Wheeler, ``Austerity Principle", China Lectures ed. L. Z. Fang
(Anhui, China, 1984)

\bibitem {Connes}
A. Connes, {\it Noncommutative Geometry} (Academic Press, New York, 1994)

\bibitem {Penrose}
R. Penrose, in {\it Quantum Theory and Beyond}, ed. T. Bastin (Cambridge 
University, Cambridge, 1971).
C. Rovelli and L. Smolin, 
Phys. Rev. D52, 5743 (1996)

\bibitem {Isham}
C. J. Isham, Class. Quant. Grav. 6, 1509 (1989); J. Math. Phys. 35, 2157 (1994);
C. J. Isham and N. Linden, J. Math. Phys. 35, 5452 (1994)

\bibitem {York}
J. W. York, Jr., Phys. Rev. D28, 2929 (1983); D31, 775 (1985); D33,
2092 (1986)

\bibitem {AndHis}
P. R. Anderson, W. A. Hiscock and D. A. Samuel, Phys. Rev. Lett. 70,
1739 (1993);
Phys. Rev. D51, 4337 (1995)

\bibitem {HPR}
B. L. Hu, Nicholas Phillips and A. Raval, in preparation

\bibitem {HRS}
B. L. Hu, A. Raval and S. Sinha, in preparation

\bibitem {CanSci}
P. Candelas and D. W. Sciama, Phys. Rev. Lett. {\bf 38}, 1372 (1977)

\bibitem {Mottola}
E. Mottola,  Phys. Rev. D33, 2136 (1986)

\bibitem {Pavon}
D. Pavon, Class. Quant. Gravity 

\end{thebibliography}
\end{document}